\documentstyle[twocolumn,prl,aps,psfig]{revtex}

\begin{document}
\draft
\preprint{IUCM95-037}

\title{Skyrmions without Sigma Models in Quantum Hall Ferromagnets}

\author{A.H. MacDonald$^{1}$, H.A. Fertig$^{2}$, and Luis Brey$^{3}$}

\address{$^{1}$ Department of Physics, Indiana University,
 Bloomington IN 47405}
\address{$^{2}$ Department of Physics and Astronomy,
University of Kentucky, Lexington KY 40506-0055}
\address{$^{3}$ Instituto de Ciencia de Materiales (CSIC),
Universidad Aut\'onoma C-12, 28049 Madrid, Spain}

\date{\today}
\maketitle

\begin{abstract}

We report on a microscopic theory of the Skyrmion states
which occur in the quantum Hall regime.  The theory is based on the
identification of Skyrmion states in this system with
zero-energy eigenstates of a hard-core model Hamiltonian.
We find that for $N_{\phi}$ orbital states in a
Landau level, a set of Skyrmions states with orbital degeneracy
$N_{\phi}-K$ and spin quantum number $S = N/2 -K$
exists for each nonnegative integer $K$.   The energetic
ordering of states with different $K$ depends on the interaction potential.

\end{abstract}


The ground state of a two-dimensional electron system (2DES)
in the strong magnetic field regime of the quantum Hall
effect is ferromagnetic at certain values of the Landau level filling
factor $\nu$.  ($\nu \equiv N/N_{\phi}$
where $N$ is the number of particles,
$N_{\phi} = A e B /h c \equiv A/ (2 \pi \ell^2)$ is the orbital
degeneracy of the Landau levels and
$\ell$ is used below as the unit of length.)
The simplest example of a quantum Hall ferromagnet (QHF)
occurs at $\nu =1 $ where the ground state is a strong
ferromagnet with total spin quantum number $S =N/2$.
Phenomena associated with {\it spontaneous}
magnetic order are open to experimental study in QHF's, despite the strong
magnetic fields, because the Zeeman coupling to the electronic spins is
smaller than other typical energy scales of the system and can
even be tuned to zero, for example by application of hydrostatic
pressure to the host semiconductor.  QHF's have the unusual property,
first identified in finite-size exact diagonalization
studies\cite{ednumerical} and dramatically evident in recent
Knight shift spin-polarization measurements\cite{barrett},
that $S$ can be sharply reduced\cite{hf},
in appropriate circumstances even to $S=0$, by the addition
or removal of a single electron.  This behavior may be
explained\cite{sondhi} using
a non-linear $\sigma$ model ($NL\sigma$) continuum field
theory description of QHF's.  In two dimensions,
the $NL\sigma$ model supports spin texture
excitations, known as Skyrmions, that carry a unit quantized topological
charge\cite{rajaraman}.
It has been conjectured\cite{sondhi,kane}, and
under appropriate circumstances it may be explicitly proven\cite{moon},
that for QHF's the {\it electrical} charge of a Skyrmion is equal to
the product of the ground state filling factor and the
topological charge.  This implies that Skyrmions
will be present in the ground state for $\nu $ close but
not equal to $1$, explaining\cite{hf} the reduction in the spatially
averaged moment.

While the $NL\sigma$ model provides
a pleasing qualitative explanation of the spin-polarization
experiments, it cannot be used to address quantitative issues.
It is valid only for slowly varying
spin-textures, while the Zeeman coupling at experimentally
relevant fields favors small Skyrmion states with only
a few reversed spins and relatively rapid variation in the spin-moment
orientation.  Recent microscopic Hartree-Fock\cite{moon,hf} estimates
of the optimal skyrmion size agree well with experiment\cite{barrett},
further bolstering evidence for unit charge,
large spin, quasiparticles of the $\nu=1$
ferromagnetic ground state.  In this Letter we report on an independent,
fully microscopic, picture of QHF Skyrmions.
In addition to giving an alternate physical picture of these
exotic quasiparticles, our approach has the advantage that we are able to
precisely determine the quantum numbers and
multiplicities of all Skyrmion states.

The approach we have taken is in the same spirit as
the illuminating outlook on the spin-polarized fractional
quantum Hall effect which arises from appropriate hard-core model
Hamiltonians\cite{haldane,macdmur}.  As discussed for the case of
interest below, these models have zero energy many-particle
eigenstates which are often known analytically, are separated from
other many-particle states by a finite gap, and have a
degeneracy which increases with increasing $N$.
An incompressible state responsible\cite{leshouches}
for a quantum Hall effect transport anomoly in such a model
is the nondegenerate maximum $N$ zero energy eigenstate.  The zero energy
eigenstates at lower densities constitute the portion of the
spectrum which involves only the degrees of freedom
of the, in general fractionally charged\cite{laughlin},
quasiholes of the incompressible state.  It is assummed that
the difference between the model Hamiltonian and the true
Hamiltonian is a sufficiently weak perturbation
that the quasihole states are still
well separated from other states in the Hilbert space, although
accidental degeneracies will be lifted in the spectrum of the
true Hamiltonian.  Here we apply this approach at $\nu =1$.
Our principle results may be summarized as follows.  The zero energy
N-fermion eigenstates for a single hole in a Landau level may be mapped
to a set of N-boson states in which the bosons are 
allowed to occupy only four single-particle states.
Single-hole states exist
with total spin number $S=N/2-K$ for each nonnegative integer $K$ 
and in the absence of disorder and Zeeman coupling  have 
degeneracy $g=g_{orb} \, g_{spin}$
where $g_{spin}=2S+1$ is the spin-multiplicity,
and the orbital degeneracy $g_{orb}=N+1-K$.

For our analysis we use the symmetric gauge in which the
single-particle orbitals\cite{leshouches} in the lowest Landau level are
\begin{equation}
\phi_m(z) = \frac{z^m}{(2^{m+1} \pi m!)^{1/2}} \exp ( - |z|^2/4),
\label{eq:1}
\end{equation}
where\cite{remarknorb} $m = 0,1, \cdots, N_{\phi}-1$ ,
$z = x+ i y$, and $x$ and $y$ are the Cartesian components of the
two-dimensional coordinate.  We study here the hard-core model for which the
interaction is\cite{remark}:
\begin{equation}
V  =  4 \pi  V_0 \sum_{i<j}  \delta^{(2)}(\vec r_i - \vec r_j)
\label{eq:ham}
\end{equation}
At strong magnetic fields the low-energy Hamiltonian is simply the
projection of this interaction onto the the lowest Landau level\cite{haldane}.
Many-particle wavefunctions which are zero energy eigenstates of this
Hamiltonian must vanish when any two-particles are at the
same position and must therefore have the
difference coordinate for each pair of particles as a factor:
\begin{equation}
\Psi[z,\chi] = \big[ \prod_{i<j} (z_i-z_j) \big] \,  \Psi_B[z,\chi].
\label{eq:map}
\end{equation}
We note that the each complex coordinate appears to the power
$N-1$ in the factor in square brackets in Eq.~(\ref{eq:map})
and that this factor is completely antisymmetric.  It follows that
$\Psi_B[z]$ must be a wavefunction for $N$ {\it bosons} and that
these bosons can be in states with angular momenta from
$0$ to $N_{\phi}-N$.  This simple observation leads to the
conclusions we reach below.

First we consider the case of a filled Landau
level, $N = N_{\phi}$.  In this case all bosons must be in
orbitals with $m=0$.  $\Psi_B[z,\chi]$ must then be
proportional to a symmetric many-particle spinor and therefore
have total spin quantum number $S=N/2$.  The orbital part of
the fermion wavefunction can be recognized as the Slater
determinant with all orbitals from $m=0$ to $m=N_{\phi}-1$ occupied.
We are able to conclude
that the ground state is a strong ferromagnet with no
orbital degeneracy.  The ease with which this conclusion can
be reached contrasts markedly with the case of the Hubbard model
where enormous effort has yielded relatively few firm results\cite{hubbard}.
When Zeeman coupling is included in the Hamiltonian the ground
state will be the member of this multiplet for which all
spins are aligned with the magnetic field, {\it i.e.}
the state with $S_z =S =N/2$.

Our primary interest here is in the elementary charged
excitations of the ferromagnetic $\nu =1$ ground state
which occur at $N=N_{\phi} \pm 1$.  For the case of a single
{\it hole} ($N = N_{\phi} - 1$) the lowest energy states are the
zero-energy eigenstates of the hard-core model.
We will see that they are the quantized version of the charged Skyrmion states
in the $NL\sigma$ model description of quantum
Hall ferromagnets.\cite{sondhi}  From Eq.~(\ref{eq:map})
it follows that these states can be mapped to spin-$1/2$ $N$ boson
states where the bosons can have angular momentum equal to
$0$ or $1$.  In understanding these states it is helpful to start
from the state with boson occupation numbers $n_{1\uparrow}=N$,
$n_{1\downarrow}=0$, $n_{0\uparrow}=0$, and $n_{0\downarrow}=0$.
Using Eq.~(\ref{eq:map}) we see that the corresponding fermion
state has fermion occupaton numbers $n_{m\uparrow}=1$ for
$m=1, \cdots, N_{\phi}$ and $n_{0\uparrow}=0$ , $n_{m\downarrow}\equiv 0$,
{\it i.e.} it is the fully polarized state with a single hole
in the $m=0$ orbital.  This state is the unique state in the
one hole Hilbert speace with the maximum possible total boson
angular momentum ($M= N$) {\it and} the maximum $S_z$ ($ = N/2$)
The set of boson states with
angular momentum $M = N - \delta M$ and $S_z = N/2 -\delta S_z$
have boson occupation numbers satisfying
\begin{eqnarray}
n_{0\downarrow}+n_{0\uparrow}&=&\delta M \nonumber\\
n_{0\downarrow}+n_{1\downarrow}&=&\delta S_z.
\label{eq:counting}
\end{eqnarray}
For fixed $\delta M$ and $\delta S_z$, the state may be
specified by $n_{0\downarrow}$, which can assume values
from $0$ to the mimimum of $\delta M$ and $\delta S_z$;
the number of states is $g(\delta S_z,\delta M) =
1 + \inf[{\delta S_z,\delta M}]$.  We now deduce the total spin quantum
numbers of the quasihole states from this expression.

Since the Hamiltonian is spin-rotationally
invariant, $S^2$ and $S_z$ must be good quantum numbers,
so that all eigenstates  occur in spin-multiplets
with degeneracy $2 S +1$ for total spin $S$.
Furthermore, assuming that
edge effects are irrelevant, the Hamiltonian is invariant
under simultaneous translation of all coordinates.  It follows
that $\delta M$ is also a good
quantum number, and that each
member of a spin-multiplet has associated with it a
large orbital degeneracy which scales with the
system size.  Orbitally degenerate states can be generated from a seed
state with minimum $\delta M$ by repeated application of the
operator which lowers the center-of-mass angular momentum\cite{trugetc}.
For example, when $S_z = N/2$ 
this procedure generates holes which occur in successively larger
single-particle angular momentum states.

\begin{figure}
\psfig{figure=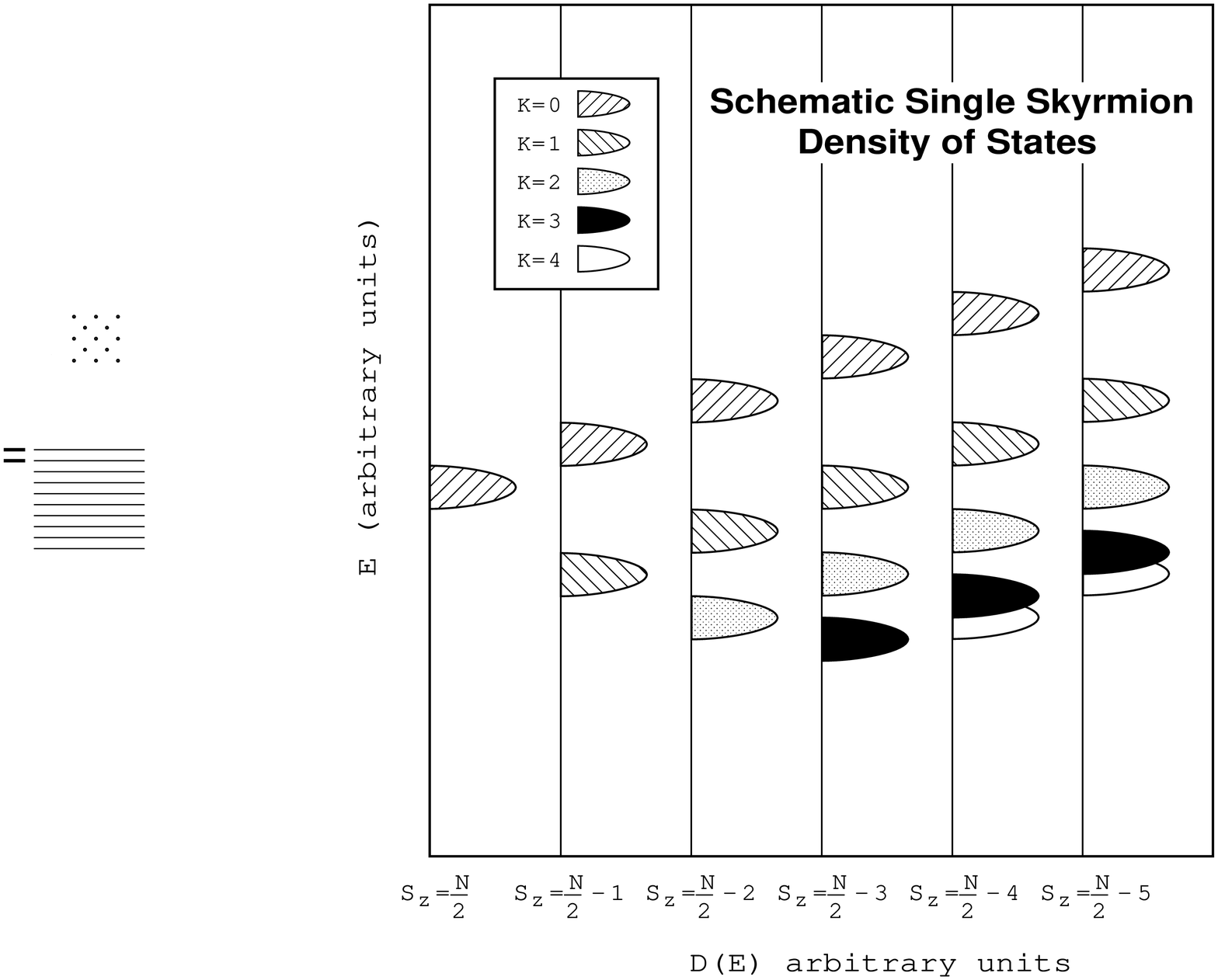,height=2.0in,width=2.0in}
\caption {Schematic single Skyrmion density of states.
For quasihole states the orbital degeneracy for each
$S=N/2-K$, $K=0,1,2,\cdots$, and $S_z=-S,\cdots,S$ is lifted by 
disorder producing a finite width band of states.  The 
spin-multiplet structure persists in the presence of disorder.
The energetic offset of
bands with the same $K$ and different $S_z$ is due to Zeeman
coupling.  For the situation illustrated, the Zeeman spin-splitting
energy is comparable to the disorder produced band width.  The dependence
of the $S_z=S$ energy on $K$ depends on the interaction Hamiltonian and 
the strength of the Zeeman coupling;
the situation illustrated where the lowest energy state occurs at
$S_z = S = N/2 -3$, is typical. The $S_z = S = N/2 -5$ band 
has been removed from this illustration for clarity.}
\label{fig1}
\end{figure}

The $S_z = N/2-1$ manifold has
one state with $\delta M=0$ and two states for each $\delta M \ge 1$.
One state at each $\delta M$ is the $S_z=N/2-1$ member of
the $S=N/2$ spin-multiplet with the same $\delta M$.  It follows that there is
one $S=N/2-1$ spin-multiplet with an orbital degeneracy which is reduced
by one compared to the $S = N/2$ states.
Continuing in the same way we may conclude
that there is a single spin-multiplet 
with $S=N/2 -K$ for each nonnegative integer $K$ with orbital degeneracy
$N_{\phi}-K$.  Thus the quantized
Skyrmion states occur in degenerate manifolds labeled by an integer $K$
and with dimension $(N_{\phi}-K)(N-2 K +1)$.
The spin degeneracy ($N-2 K + 1$) is lifted by the Zeeman
coupling and is the quantum counterpart of the arbitrary global orientation
of a classical Skyrmion.  The orbital degeneracy ($N_{\phi}-K$) is
the quantum counterpart of the arbitrary location of the center of a
classical Skyrmion and is lifted by a disorder potential.  The density of
Skyrmion states in the presence of Zeeman coupling
and weak disorder is indicated schematically in Fig.~\ref{fig1}.

Repeated application of the total spin-lowering operator and
the center-of-mass lowering operator allows all states in the
$S=N/2-K$ manifold to be generated from
the seed state which has $S_z= S $ and $\delta M = K$.
To determine the many-body wavefunction of this spin-$K$ Skyrmion
state we consider the effect of a total spin raising operator, $S_{+}$, on the
$K+1$ boson states which occur at $\delta M = \delta S_z =K$:
\begin{eqnarray}
& & S_{+} |n_{0\downarrow} = k,n_{0\uparrow} = K-k,n_{1\downarrow} = K-k
\rangle\label{eq:splus} \\
 &=&  \sum_{m=0}^{1} b^{\dagger}_{m\uparrow} b_{m\downarrow} |n_{0\downarrow} =
k,n_{0\uparrow}
 = K-k,n_{1\downarrow} = K-k \rangle \nonumber \\
 &\approx& \sqrt{(K-k) N } | n_{0\downarrow} = k, n_{0\uparrow}
 = K-k,n_{1\downarrow} = K-k-1 \rangle. \nonumber
\end{eqnarray}
Here $b^{\dagger}_m$ and $b_m$ are boson creation and annihilation operators.
In the last form of Eq.~(\ref{eq:splus}) we have included only the
$m=1$ term which dominates for finite $K$ and $N \to \infty$ because 
$\sqrt{n_{1\uparrow}} \approx \sqrt{N} \gg 1$.  In this limit
the state with $k=K$ is annihilated by $S_{+}$ and is therefore
the seed state of the $S=N/2-K$ Skyrmion multiplet.  In first
quantization the (unnormalized) boson wavefunction of this state is
\begin{eqnarray}
|\Psi^{SK}_K \rangle &=& \sum_{i_1,\cdots,i_K}^{\prime}
\big[ \prod_{j \in \{i_K\}} |\downarrow\rangle_j \exp ( -|z_j|^2/4 )
\big]\nonumber\\
 && \big[ \prod_{l \ni \{i_K\}} |\uparrow\rangle_l z_l \exp ( -|z_l|^2/4)\big]
\label{eq:skwf}
\end{eqnarray}
In these wavefunctions, the sums are over distinct particle indices,
up-spin quasiparticles occupy states with angular momentum $m=1$ whereas
down-spin particles occupy $m=0$ states;
this correlation between spin and angular momentum states is an essential
aspect of Skyrmion states\cite{hf,moon}.

We now establish some relationships
between the properties of these wavefunctions and previously known
results.  As explained above the $K=0$ Skyrmion wavefunction
is identical to a Hartree-Fock quasihole.  On the other hand, the
large $K$ limit of the these wavefunctions
may be related to classical field theory Skyrmions.
Consider the fermion wavefunction
\begin{eqnarray}
|\Psi(\lambda)\rangle &=& \sum_{K=0}^{\infty} \lambda^K |\Psi^{SK}_K
\rangle\nonumber\\
 &=& \prod_{ i < j} (z_i-z_j) \prod_{l}(z_l |\uparrow\rangle_l +
\lambda |\downarrow\rangle_l)\nonumber\\
 && \exp(-|z_l|^2/4).
\label{eq:psilam}
\end{eqnarray}
For large $\lambda$ the sum in Eq.~(\ref{eq:psilam}) will be
dominated by terms in a relatively narrow range of $K$
values\cite{notquitetrue}.
$|\Psi(\lambda)\rangle$ is precisely the single-Slater-determinant
proposed by Moon {\it et al.}\cite{moon} as a microscopic trial
wavefunction for the Skyrmion on the grounds that, for large $\lambda$,
it gives a spin-texture in precise agreement with
that of the classical Skyrmion\cite{sondhi} of size $\lambda$.

The more general Hartree-Fock (HF) single-Slater-determinant
wavefunctions of Ref.~\cite{hf} may be written in the form
\begin{equation}
|\Psi^{HF}\rangle = \prod_m\bigl( u_m c_{m\downarrow}^{\dag}
+v_m c_{m+1\uparrow}^{\dag} \bigr) |0\rangle.
\label{eq:hfwf}
\end{equation}
The HF approximation is equivalent to minimizing the
energy of this wavefunction subject to a normalization constraint.
An obvious deficiency of the Hartree-Fock approximation is its
failure to reflect known symmetries of the Hamiltonian.  In particular,
the Hartree-Fock wavefunction is not an eigenstate of $S_z$ or $M$.
As discussed by Nayak and Wilczek\cite{nayak},
this failure is readily remedied by projecting the Hartree-Fock
state onto a state of definite $S_z$ (and therefore definite $M$).
For the hard-core model, the HF equations\cite{hf}
may be solved {\it analytically}, with the result
$|u_m|^2=1-|v_m|^2=\lambda^2/[\lambda^2+ 2 (m+1)]$, where $\lambda$ is
a free parameter.  It is easily verified that the corresponding
wavefunction is precisely $|\Psi(\lambda)\rangle$.
In the case of the hard-core model the projection of the
Hartree-Fock wavefunction onto a state of definite $S_z$
yields the exact Skyrmion wavefunctions, suggesting that
this seemingly {\it ad hoc} procedure might be generically accurate.

We remark that all the results we have obtained here to  
a hole in the ferromagnetic ground state apply equally well
to the case of a particle added to the ferromagnetic ground
state because of an exact particle-hole symmetry\cite{phsym} which
holds around $\nu =1$.  For the case of the hard-core model
the energy of the particle states is $4 V_0$, independent of $K$.
Our analysis is also readily generalized for fractional
filling factors $\nu =1/m$, although in that case the quasiparticle
states cannot be generated by particle-hole transformation.
For general interactions the Skyrmion energy is dependent
on $K$; the minimum energy state may occur at $K =0$
where Hartree-Fock theory is valid, at $K \to \infty$
where the classical field theory becomes valid, or
at an intermediate value of $K$ where explicit expressions
for the energy are not available.  The many-particle
wavefunctions which we have derived for the Skyrmion states
are exact only for the hard-core model.  The accuracy of
these wavefunctions for general interactions, which has 
been confirmed by exact diagonaliztion\cite{juanjo}, rests on
the existence of a gap for charged excitations and they are
,in this sense, analogous to Laughlin's trial wavefunctions for
incompressible states at fractional filling factors.
This work was supported in part by  NATO Collaborative Research Grant No.
930684, by the National Science Foundation under grants DMR-9416906,
and DMR-9503814 and by CICyT of Spain under contract MAT 94-16906.
HAF acknowledges the support of the A.P. Sloan Foundation
and the Research Corporation.  Helpful conversations with Steve Girvin,
Kyungsun Moon, Kun Yang, Carlos Tejedor, Juan-Jose Palacios, and Luis
Martin-Moreno are gratefully acknowledged.

\begin{references}
\bibitem{notquitetrue} There is, however, a tail in the probability
distribution of $K$ values in $|\Psi(\lambda)\rangle$
which causes $\langle K \rangle$ to diverge logarithmically for $N \to \infty$.
\end{references}
\end{document}